\documentclass[article,showpacs,preprintnumbers,amsmath,amssymb]{revtex4}
\usepackage[dvips]{graphicx}
\usepackage{dcolumn}
\usepackage{bm}    

\usepackage{graphicx,amsmath,bm}

\begin{document}

\title{Chaos in three coupled rotators:\\From Anosov dynamics to hyperbolic
attractors}


\author{Sergey P. Kuznetsov\textsuperscript{1,2}}
\affiliation{\textsuperscript{1} Institute of Computer Science, Udmurt State University, Universitetskaya 1, Izhevsk, 426034, Russia\\ \\
\textsuperscript{2} Kotelnikov's Institute of Radio-Engineering and Electronics of RAS,
Saratov Branch, Zelenaya 38, Saratov, 410019, Russia}

\begin{abstract}
Starting from Anosov chaotic dynamics of geodesic flow on a surface of negative curvature, we develop and consider a number of self-oscillatory
systems including those with hinged mechanical coupling of three rotators
and a system of rotators interacting through a potential function. These
results are used to design an electronic circuit for generation of rough
(structurally stable) chaos. Results of numerical integration of the model
equations of different degree of accuracy are presented and discussed. Also, circuit simulation of the
electronic generator is provided using the NI Multisim environment.
Portraits of attractors, waveforms of generated oscillations,
Lyapunov exponents, and spectra are considered and found to be in good
correspondence for the dynamics on the attractive sets of the
self-oscillatory systems and for the original Anosov geodesic flow. The
hyperbolic nature of the dynamics is tested numerically using a criterion
based on statistics of angles of intersection of stable and unstable
subspaces of the perturbation vectors at a reference phase trajectory on the
attractor.
\end{abstract}



\maketitle

\section*{Introduction}

Hyperbolic theory is a branch of the theory of dynamical systems, which
provides a rigorous justification for chaotic behavior of deterministic
systems with discrete time (iterative maps -- diffeomorphisms) and with
continuous time (flows) \cite{01,02,03,04,05}. Objects of consideration in this framework
are uniformly hyperbolic invariant sets in phase space, composed
exclusively of saddle trajectories. For conservative systems, hyperbolic
chaos is represented by Anosov dynamics, when a uniformly hyperbolic
invariant set occupies completely a compact phase space (for a
diffeomorphism) or a constant energy surface (for a flow). For dissipative
systems, the hyperbolic theory introduces a special type of attracting
invariant sets, the uniformly hyperbolic chaotic attractors.

A fundamental mathematical fact is that uniformly hyperbolic invariant sets
possess a property of roughness, or structural stability. It means that with
small variations (perturbations) of the system, the character of the
dynamics is preserved up to a continuous variable change.

After Andronov and Pontryagin introduced the concept of roughness applicable
originally to systems with regular dynamics \cite{06}, it is commonly used in the
theory of oscillations to postulate that the rough systems are of primary
theoretical and practical interest because of their insensitivity to
variations in parameters, manufacturing errors, interferences, etc. \cite{07,08,09}.
This proposition seems general and convincing. Therefore, in the context of
complex dynamics, one could expect that the hyperbolic chaos as rough
phenomenon should occur in many physical situations. Moreover, just these
systems should be of preferring interest for chaos applications \cite{10,11,12}.

Paradoxically, consideration of numerous examples of complex dynamics
related to different fields of science does not justify the expectations
regarding prevalence of the hyperbolic chaos. Anosov once said that \textit{``One gets the impression that the Lord God would prefer to weaken hyperbolicity a bit rather than deal with the restrictions on the topology of an attractor that arise when it really is (completely and uniformly) {\lq 1960s-model\rq}  hyperbolic''} \cite{13}.
Therefore, the scientific community began to consider the hyperbolic
dynamics only as a refined abstract image of chaos, while the efforts of
mathematicians were redirected to the development of more widely applicable
generalizations \cite{14,15}.

In this situation, instead of searching for "ready-to-use" examples from
real world, it makes sense to turn to a purposeful construction of systems
with hyperbolic chaos basing on tools of physics and technology. For this
purpose it is natural to exploit the property of roughness or structural
stability \cite{16,17}; namely, taking as a prototype some formal example of
hyperbolic dynamics, one can try to modify it in such way that the dynamic
equations would correspond as far as possible to a physically implementable
system. Due to the roughness, it may be hoped that the hyperbolic chaos will
retain its nature under the transformation from a formal model towards a
realizable system. The principal point is that at each step of the
construction process important is monitoring the hyperbolic nature of the
dynamics. It is difficult to expect that systems designed on physical basis
will be convenient for rigorous mathematical proofs; so, it is natural
to turn to methods of computer testing of the hyperbolicity.

An idea of verifying hyperbolicity is based on computational analysis of
statistics of angles between stable and unstable subspaces nearby a reference
phase trajectory as proposed originally in \cite{36} for saddle invariant sets.
Subsequently, this method was developed and used as well to chaotic
attractors \cite{37,38,39,40,41,42,16,17}.

The technique consists in the following. Along a typical phase trajectory
belonging to the invariant set of interest, we follow it in forward time and
in reverse time, and evaluate at each point an angle between the subspaces
of perturbation vectors to analyze their statistical distribution. If the
resulting distribution does not contain angles close to zero, it indicates a
hyperbolic nature of the invariant set. If positive probability for zero
angles is found, the tangencies between the stable and unstable manifolds of
the trajectory do occur, and the invariant set is not hyperbolic. The latter
may indicate the presence of a quasi-attractor that is a complex set, which
contains long-period stable cycles with very narrow domains of attraction \cite{02}.

In this paper, starting with a classic problem of geodesic flow on a surface
of negative curvature, we elaborate several modifications of the mechanical
hinge system in variants exhibiting chaotic self-oscillations \cite{27}, and by
analogy with them design an electronic device operating as a generator of
rough chaos \cite{49,50}.

\subsection{The geodesic flow on the surface of negative curvature and the mechanical model}

It is known that a free mechanical motion of a particle in a space with
curvature is carried out along the geodesic lines of the metric associated
with a quadratic form expressing the kinetic energy $W$ in terms of generalized
velocities with coordinate-dependent coefficients \cite{18,19,20}. In particular, in
the two-dimensional case, using coefficients of the quadratic form
\begin{equation}
\label{eq1}
W = E(x,y)\dot {x}^2 + 2F(x,y)dxdy + G(x,y)dy^2,
\end{equation}
one can find the curvature with the formula of Gauss -- Brioschi
known in differential geometry \cite{21,22}:
\begin{equation}
\label{eq2}
\begin{array}{c}
 K = (A - B)/(EG - F^2)^{2}, \\ \\
 A = \left| {{\begin{array}{*{20}c}
 { - \textstyle{1 \over 2}E_{yy} + F_{xy} - \textstyle{1 \over 2}G_{xx} }
\hfill & {\textstyle{1 \over 2}E_x } \hfill & {F_x - \textstyle{1 \over
2}E_y } \hfill \\
 {F_y - \textstyle{1 \over 2}G_x } \hfill & E \hfill & F \hfill \\
 {\textstyle{1 \over 2}G_y } \hfill & F \hfill & G \hfill \\
\end{array} }} \right|, \,\,\,
 B = \left| {{\begin{array}{*{20}c}
 0 \hfill & {\textstyle{1 \over 2}E_y } \hfill & {\textstyle{1 \over 2}G_x }
\hfill \\
 {\textstyle{1 \over 2}E_y } \hfill & E \hfill & F \hfill \\
 {\textstyle{1 \over 2}G_x } \hfill & F \hfill & G \hfill \\
\end{array} }} \right|, \\
 \end{array}
\end{equation}
where the subscripts denote the corresponding partial derivatives. In the
case of negative curvature, the motion is characterized by instability with
respect to transverse perturbations. Therefore, if it occurs in a bounded
region, it turns out to be chaotic \cite{19, 20}.

\subsubsection{The Anosov geodesic flow on the Schwartz surface}

As a concrete example, consider a geodesic flow on the so-called
\textit{minimal Schwartz P-surface} \cite{23}. In
the space $(\theta _1 ,\,\,\theta _2 ,\,\,\theta _3 )$ this surface is given
by the equation
\begin{equation}
\label{eq3}
\cos \theta _1 + \cos \theta _2 + \cos \theta _3 = 0.
\end{equation}
Due to periodicity along the three coordinate axes, we can assume that the
variables $\theta _{1,2,3}$ are defined modulo 2$\pi $, and to interpret
the motion as occurring in a compact region, namely, a cubic cell of edge
length 2$\pi $.

We assume that conservative dynamics on the surface (\ref{eq3}) proceed with
conservation of the kinetic energy
\begin{equation}
\label{eq4}
W = \textstyle{1 \over 2}(\dot {\theta }_1^2 + \dot {\theta }_2^2 + \dot
{\theta }_3^2 ),
\end{equation}
where the mass is taken as unity, and the relation (\ref{eq3}) corresponds to a
constraint imposed on the system. Expressing one of the generalized
velocities in terms of the other two, we obtain
\begin{equation}
\label{eq5}
W = \textstyle{1 \over 2}\left[ {P(\theta _1 ,\theta _2 )\dot {\theta }_1^2
+ 2Q(\theta _1 ,\theta _2 )\dot {\theta }_1 \dot {\theta }_2 + R(\theta _1
,\theta _2 )\dot {\theta }_2^2 } \right],
\end{equation}
where
\begin{equation}
\label{eq6}
\begin{array}{l}
 P(\theta _1 ,\theta _2 ) = 1 + \frac{\sin ^2\theta _1 }{1 - (\cos \theta _1 + \cos \theta _2
)^2},\,\,\,\, \\ \\
 Q(\theta _1 ,\theta _2 ) = \frac{\sin \theta _1 \sin \theta _2 }{1 - (\cos \theta _1 + \cos \theta
_2 )^2},\,\,\, \\ \\
 R(\theta _1 ,\theta _2 ) = 1 + \frac{\sin ^2\theta _2 }{1 - (\cos \theta _1 + \cos \theta _2 )^2}.
\end{array}
\end{equation}

Completely, the metric is defined by (\ref{eq5}) and (\ref{eq6}) supplemented by formulas
for other sheets of the ``atlas of consistent maps'' \cite{28} of the
two-dimensional manifold, which are obtained by cyclic permutation of the
indices.

The Gauss--Brioschi formula for curvature in this case leads to explicit
expression, which, taking into account the constraint equation (\ref{eq3}), can be
rewritten in a symmetric form \cite{25,26}:
\begin{equation}
\label{eq7}
K = - \frac{\cos ^2\theta _1 + \cos ^2\theta _2 + \cos ^2\theta _3 }{2(\sin
^2\theta _1 + \sin ^2\theta _2 + \sin ^2\theta _3 )^2}.
\end{equation}

With exception of eight points, where the numerator vanishes because all
three cosines are equal to zero, the curvature $K$ is negative, so
that the geodesic flow is Anosov's.

Using the standard procedure for mechanical systems with holonomic
constraints \cite{30, 31}, we can write down the set of equations of motion in
the form
\begin{equation}
\label{eq8}
\ddot {\theta }_1 = - \Lambda \sin \theta _1 ,\,\,\ddot {\theta }_2 = -
\Lambda \sin \theta _2 ,\,\,\ddot {\theta }_3 = - \Lambda \sin \theta _3 ,
\end{equation}
where the Lagrange multiplier $\Lambda $ has to be determined taking into
account the algebraic condition of the mechanical constraint, which
supplements the differential equations. In our case
\begin{equation}
\label{eq9}
\Lambda = - \frac{\dot {\theta }_1^2 \cos \theta _1 + \dot {\theta }_2^2
\cos \theta _2 + \dot {\theta }_3^2 \cos \theta _3 }{\sin ^2\theta _1 + \sin
^2\theta _2 + \sin ^2\theta _3 }.
\end{equation}

The system (\ref{eq8}) has first integrals, one of which corresponds to the
constraint equation (\ref{eq3}), and the other to its time derivative, so the
dimension of the phase space is reduced to four. In addition, there is an
energy integral due to the conservative nature of the dynamics.

Figure 1 shows a trajectory in the configuration space
obtained from numerical integration of the equations.
When plotting the picture, the angular variables are related to the interval
from 0 to 2$\pi $, i.e., the diagram in the three-dimensional space ($\theta
_{1}$,$\theta _{2}$,$\theta _{3})$ corresponds to a single fundamental
cubic cell, which reproduces itself periodically with a shift by 2\textit{$\pi $} along each
of three coordinate axes. The points are located on a two-dimensional
surface, given by the equation $\cos \theta _1 + \cos \theta _2 + \cos
\theta _3 = 0$, where the mechanical constraint is fulfilled. The opposite
faces of the cubic cell naturally may be identified; as a result we arrive
at a compact manifold of genus 3 that is a surface topologically equivalent
to a "pretzel with three holes" \cite{24, 25}. From the picture we can conclude
visually about chaotic nature of the trajectory, which covers the surface
ergodically. The power spectrum of the signal generated by the motion of the
system is continuous that corresponds to chaos (Fig. 2).
\begin{figure}[!t]
\centering{
\includegraphics[width=6cm]{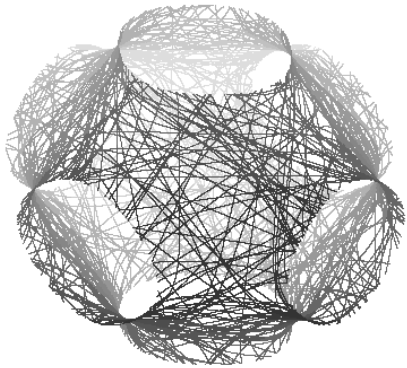}
}
\caption{A typical trajectory in the three-dimensional configuration space
($\theta _{1}$, $\theta _{2}$, $\theta _{3})$ of the system (\ref{eq8}), (\ref{eq9})}\label{fig1}
\end{figure}
\begin{figure}[htbp]
\centering{
\includegraphics[width=9cm]{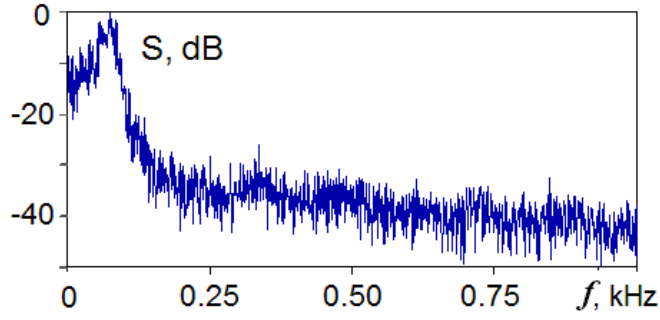}
}
\caption{The power spectrum calculated for the variable $\cos \theta _1 $
of the system (\ref{eq8}), (\ref{eq9}) for the motion with kinetic energy $W$=0.3}
\label{fig2}
\end{figure}

Taking into account the imposed mechanical constraint, there are four
Lyapunov exponents characterizing perturbations near the reference phase
trajectory: one positive, one negative and two zero ones. One exponent
vanishes due to the autonomous nature of the system; it is responsible for a
perturbation directed along the phase trajectory. The other
is associated with a perturbation of the energy shift.

Since there is no characteristic time scale in the system, the Lyapunov
exponents corresponding to the exponential growth and decrease of
perturbations per unit time must be proportional to the velocity, i.e.
$\lambda = \pm \kappa \sqrt W $, where the coefficient is determined by the
average curvature of the metric. Empirically, for the system under
consideration the calculations yield $\kappa = 0.70$ \cite{26, 27}.

\subsubsection{The triple linkage}
Dynamics corresponding to the geodesic flow on the Schwartz surface takes
place in the Thurston--Weeks--MacKay--Hunt triple hinge mechanism \cite{24,25}
shown schematically in Fig. 3

The Cartesian coordinates of the hinges P$_{1,2,3}$ attached to the disks
are expressed through the angles $\theta _{1}$, $\theta _{2}$, $\theta
_{3}$, counted from the rays connecting the disk centers with the origin
as follows:
\begin{equation}
\label{eq10}
\begin{array}{l}
 x_1 + iy_1 = 1 - re^{i\theta _1 },\,\, \\
 x_2 + iy_2 = - e^{2\pi i / 3}(1 - re^{i\theta _2 }),\,\,\,\, \\
 x_3 + i\,y_3 = - e^{ - 2\pi i / 3}(1 - re^{i\theta _3 }). \\
 \end{array}
\end{equation}

The mechanical constraint provided by the rods and hinges implies that the
radius of the circumscribed circle of the triangle P$_{1}$P$_{2}$P$_{3}$
must be $R$. By the known formula, $R = abc / 4S$, where $a$, $b$, $c$ are sides of
the triangle, and $S$ is its area. With given coordinates of three vertices
($x_{i}$, $y_{i})$, we evaluate the sides, and the area is expressed in terms
of the vector product of vectors ${\rm {\bf b}} = P_1 P_2 $ and ${\rm {\bf
c}} = P_1 P_3 $, $S = \textstyle{1 \over 2}\left| {{\rm {\bf b}}\times {\rm
{\bf c}}} \right|$. Thus, we have $(abc)^2 - 4R^2({\rm {\bf b}}\times {\rm
{\bf c}}) = 0$, or
\begin{equation}
\label{eq11}
\begin{array}{c}
 [(x_1 - x_2 )^2 + (y_1 - y_2 )^2]
 \cdot [(x_2 - x_3 )^2 + (y_2 - y_3 )^2]
 \cdot [(x_3 - x_1 )^2 + (y_3 - y_1 )^2] / 4R^2 \\
 - (x_2 y_3 + x_3 y_1 + x_1 y_2 - x_3 y_2 - x_1 y_3 - x_2 y_1 )^2 = 0. \\
 \end{array}
 \end{equation}
Substituting the expressions from (\ref{eq10}), we obtain the constraint equation in
the form $F(\theta _1 ,\theta _2 ,\theta _3 ) = 0$.
\begin{figure}[!t]
\centering{
\includegraphics[width=6cm]{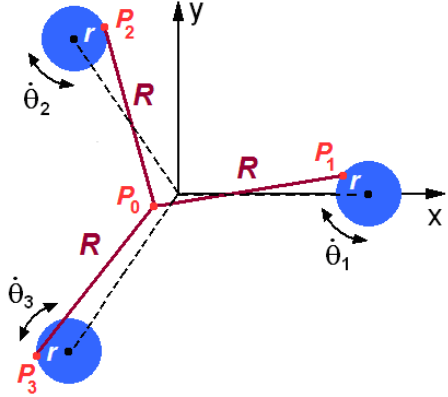}
}
\caption{The triple linkage of Thurston--Weeks--MacKay--Hunt \cite{24,25} is made
of three disks placed in a common plane, with centers at the vertices of
an equilateral triangle, each of which is capable of rotating about its
axis. On each disk at the edge a hinge is attached (P$_{1,2,3}$), and three
identical rods are connected to these hinges, the opposite ends of which are
joined together by another movable hinge (P$_{0}$)}\label{fig3}
\end{figure}

Assuming $r < < 1$ and expanding (\ref{eq11}) in Taylor series up to terms of the
first order in the small parameter, we obtain
\begin{equation}
\label{eq12}
27(R^2 - 1) - 18r(2R^2 - 3)(\cos \theta _1 + \cos \theta _2 + \cos \theta _3
) = 0.
\end{equation}

With $R = 1$ we arrive at the constraint equation, which is exactly the same
as used in the context of the geodesic flow on the Schwarz surface (\ref{eq3}).
Under assumption that the only massive elements are the
disks, with unit moment of inertia, the equations of motion are reduced
exactly to the form (\ref{eq8}).

\subsection{Self-oscillating systems with mechanical hinge connections}

Suppose that additional torques $M_{1,2,3}$ are applied to the disks of
the triple linkage system, then, in the assumptions we used, the equations are \cite{26,27}
\begin{equation}
\label{eq13}
\ddot {\theta }_1 = M_1 - \Lambda \sin \theta _1 ,\,\,\ddot {\theta }_2 =
M_2 - \Lambda \sin \theta _2 ,\,\,\ddot {\theta }_3 = M_3 - \Lambda \sin
\theta _3 ,
\end{equation}
where
\begin{equation}
\label{eq14}
\Lambda = - \frac{\sum\nolimits_{j = 1}^3 {\left( {\dot {\theta }_j^2 \cos
\theta _j + M_j \sin \theta _j } \right)} }{\sum\nolimits_{j = 1}^3 {\sin
^2\theta _j } }.
\end{equation}

Setting the torques as functions of the angular velocities, one can obtain
various variants of self-oscillating systems demonstrating dynamics of the
Anosov geodesic flow on their attracting sets.

\subsubsection{System with invariant energy surface}

Consider first the case simplest for analysis as proposed
qualitatively in \cite{25}. Namely, we set
\begin{equation}
\label{eq15}
M_i = \left( \mu - \nu \sum\nolimits_{j = 1}^3 {\dot {\theta }_j^2 } \right) \dot {\theta }_i ,\,\,i = 1,2,3,
\end{equation}
where $\mu $, $\nu $ are constants. In practice, to obtain such a function,
the mechanism should be supplemented with a controlling device and
actuators, which apply the torques $M_{1,2,3}$ to the axles of the disks
depending on the value of the detected instant kinetic energy.

Equations (\ref{eq8}) in this case read
\begin{equation}
\label{eq16}
\begin{array}{l}
 \ddot {\theta }_i = \left( \mu - \nu \sum\nolimits_{j = 1}^3 {\dot {\theta }_j^2 } \right)
 \left( {\dot {\theta }_i - \frac{\sum\nolimits_{j =1}^3 {\dot {\theta }_j \sin \theta _j \sin \theta _i} }{\sum\nolimits_{j = 1}^3 {\sin ^2\theta _j } }} \right)
 - \frac{\sum\nolimits_{j = 1}^3 {\dot {\theta }_j^2 \cos \theta _j \sin \theta _i}
}{\sum\nolimits_{j = 1}^3 {\sin ^2\theta _j } }, \\ i =1,2,3. \\
 \end{array}
\end{equation}
From this it is easy to derive an equation describing evolution of the
kinetic energy $W = \textstyle{1 \over 2}(\dot {\theta }_1^2 + \dot {\theta
}_2^2 + \dot {\theta }_3^2 )$, namely,
\begin{equation}
\label{eq17}
\dot {W} = 2(\mu - 2\nu W)W.
\end{equation}

Figure 4 shows the angular velocities of the disks and the energy versus
time, in the course of the transient process of arising chaotic
self-oscillations when starting with small initial velocity from a point in the configuration
space admissible by the mechanical constraint.
As a result, the self-oscillatory regime develops corresponding to a
constant value of the kinetic energy, accompanied by irregular oscillations
of the angular velocities evidently of chaotic nature (no repetitions of
the waveforms are visible).
\begin{figure}[!t]
\centering\includegraphics[width=0.8\textwidth]{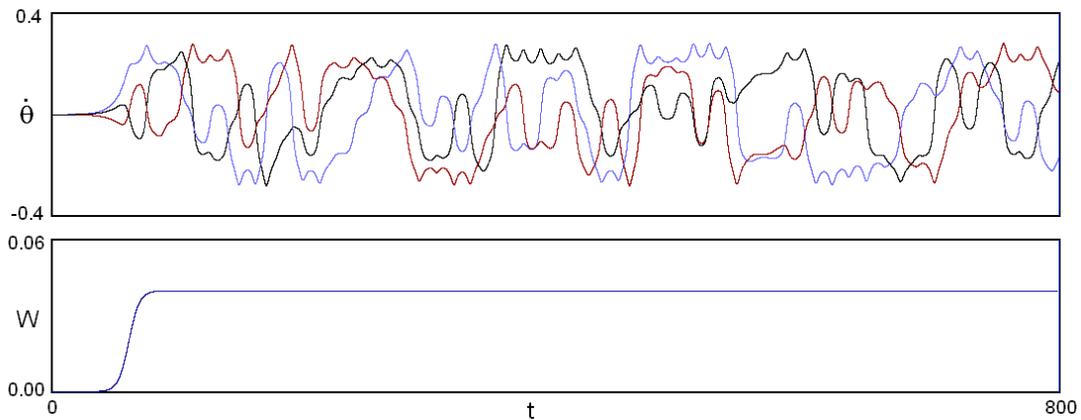}
\caption{Evolution of the angular velocities $\dot {\theta }_1 ,\,\,\dot
{\theta }_2 ,\,\,\dot {\theta }_3 $ and energy $W$ in time in the course of
transient process towards chaotic self-oscillations at $\mu $=0.12, $\nu$=1.5.}
\end{figure}

Formally, the phase space is six-dimensional, and
there are six Lyapunov exponents characterizing behavior of perturbed phase
trajectories near the reference orbit. Excluding two nonphysical zero
exponents, which violate the constraint equation, we have four exponents in
the rest. Since the motion on the attractor takes place on the energy
surface, the exponents for perturbations without departure from this surface
will be equal to those in the conservative system at the same energy:
$\kappa \sqrt W $, 0, and $ - \kappa \sqrt W $, $\kappa = 0.70$ \cite{26,27}. The
exponent corresponding to perturbation of the energy is evaluated easily as
the Lyapunov exponent of the attracting fixed point $W=\mu $/2$\nu $ in Eq.
(\ref{eq17}); it is equal to  $-2\mu $.

The observed attractor is undoubtedly hyperbolic since the dynamics takes
place along geodesic lines of the metric of negative curvature on the energy
surface, which is itself an attractive set. It is interesting, however, to
try application of the criterion of angles in this case to test the
methodology, which further will be exploited in situations where the
presence or absence of hyperbolicity is not trivial.

The procedure begins with calculation of a reference orbit ${\rm {\bf
x}}(t)$ on the attractor, for which numerical integration of the equations,
briefly written in the form
${\rm {\bf \dot {x}}} = {\rm {\bf F}}({\rm {\bf x}},t)$,
is performed over a sufficiently long time interval. Being
interested in the one-dimensional subspace associated with the largest
Lyapunov exponent, we integrate the linearized equation for the perturbation
vector along the reference trajectory: ${\rm {\bf \dot {\tilde {x}}}} = {\rm
{\bf {F}'}}({\rm {\bf x}}(t),t){\rm {\bf \tilde {x}}}$. Normalizing the
vectors ${\rm {\bf \tilde {x}}}$ at each step $n$, we obtain a set of unit
vectors $\,\{{\rm {\bf x}}_n \}$. Next, we integrate a conjugate linear
equation
${\rm {\bf \dot {u}}} = - [{\rm {\bf {F}'}}({\rm {\bf x}}(t),t)]^{T}{\rm {\bf u}}$,
where T denotes transpose, in inverse
time along the same reference trajectory \cite{41}. Then, we obtain a set of
vectors $\,\{{\rm {\bf u}}_n \}$ normalized to unity that determine the
orthogonal complement to the sum of the stable and neutral subspaces of the
perturbation vectors at the reference trajectory. Now, to evaluate an angle
$\phi $ between the subspaces at each $n$-th step, we calculate the angle
$\beta _{n} \in $[0,$\pi $/2] between the vectors ${\rm {\bf \tilde
{x}}}_n $, ${\rm {\bf u}}_n $, and set $\phi _n = \pi \mathord{\left/
{\vphantom {\pi 2}} \right. \kern-\nulldelimiterspace} 2 - \beta _n $.

Figure 5 shows histograms of the distributions of angles between stable
and unstable subspaces numerically obtained for the system (\ref{eq16}) with $\nu
$=1.5 at $\mu $=0.12 and 0.75. As seen, the distributions are clearly
distant from zero angles; so, the test confirms the hyperbolicity of the
attractors.
\begin{figure}
\centering\includegraphics[width=6cm]{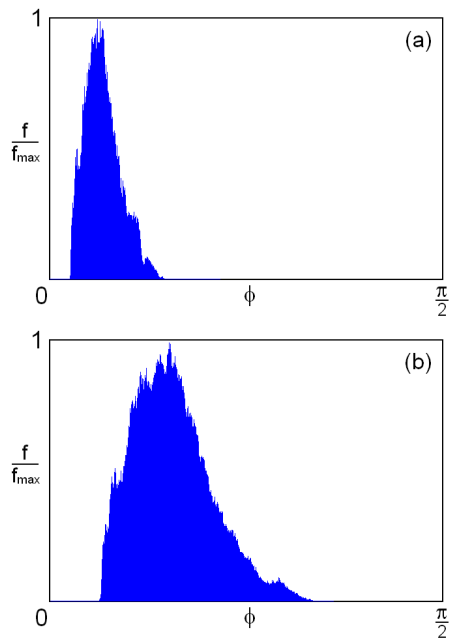}
\caption{Verification of the hyperbolicity criterion for angles in a system
(\ref{eq16}) with invariant energy surface for $\mu $=0.12 and $\mu $=0.75 at $\nu
$=1.5.}
\label{fig5}
\end{figure}

\subsubsection{System of three self-rotators with hinge constraint}

Now let us turn to a system, where a torque applied to each disk depends
only on the angular velocity of that disk. Namely, we set
\begin{equation}
\label{eq18}
M_i = (\mu - \nu \dot {\theta }_1^2 )\dot {\theta }_i ,\,\,i = 1,2,3.
\end{equation}

To provide such torques in a real device, one can use a pair of friction
clutches attached to each disk, which transmit oppositely directed
rotations, selecting properly the functional dependence of the friction
coefficient on the velocity. In this case, each of the three disks is a
self-rotator that means a subsystem, which, being singled out, manifests
evolution in time with approach to a steady rotation with constant angular
velocity $\dot {\theta } = \pm \sqrt {\mu \nu ^{ - 1}} $ in one or other
direction (depending on initial conditions).

Equations (\ref{eq3}) in this case will be rewritten in the form
\begin{equation}
\label{eq19}
 \ddot {\theta }_i = (\mu - \nu \dot {\theta }^2)\left( {\dot {\theta }_i -
\frac{\sum\nolimits_{j = 1}^3 {\dot {\theta }_j \sin \theta _j }
}{\sum\nolimits_{j = 1}^3 {\sin ^2\theta _j } }\sin \theta _i } \right)
 - \frac{\sum\nolimits_{j = 1}^3 {\dot {\theta }_j^2 \cos \theta _j }
}{\sum\nolimits_{j = 1}^3 {\sin ^2\theta _j } }\sin \theta _i ,\,\,i =
1,2,3.
\end{equation}

Figure 6 shows the angular velocities of the disks and the energy $W =
\textstyle{1 \over 2}(\dot {\theta }_1^2 + \dot {\theta }_2^2 + \dot {\theta
}_3^2 )$ versus time in the transient process towards chaotic
self-oscillations at $\mu $=0.06, $\nu $=1.5 starting with small
initial velocity from a point in the
configuration space admissible by the mechanical constraint. As a result of the transient process, a chaotic
self-oscillatory regime arises, in which the kinetic energy fluctuates
irregularly about a certain mean value. Figure 7 shows how the mean kinetic
energy and its standard deviation depend on the supercriticality $\mu$.
\begin{figure}
\centering\includegraphics[width=.8\textwidth]{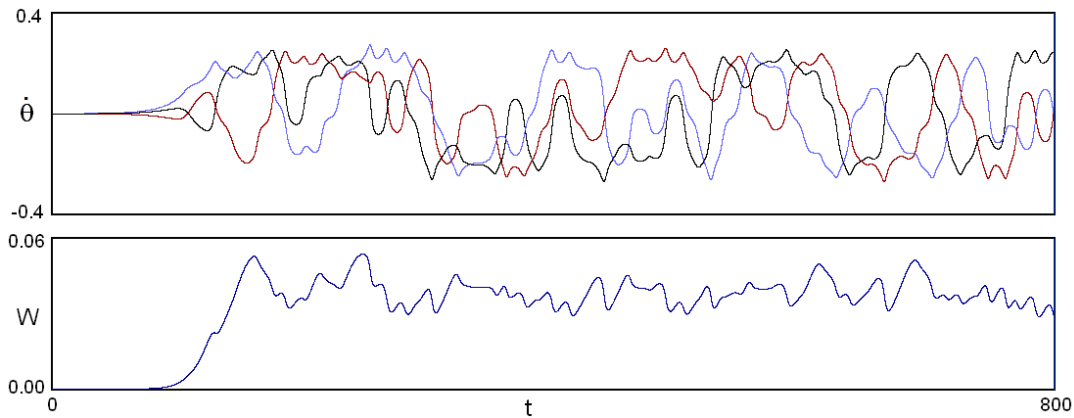}
\caption{Angular velocities $\dot {\theta }_1 ,\,\,\dot {\theta }_2 ,\,\,\dot
{\theta }_3 $ and energy $W$ versus time in the transient process towards
chaotic self-oscillations in a system of three self-rotators with mechanical
constraint (\ref{eq19}) at $\mu $=0.06, $\nu $=1.5.}
\end{figure}

\begin{figure}[!t]
\centering{
\includegraphics[width=6cm]{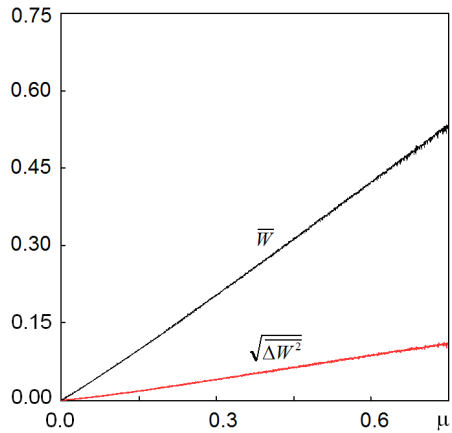}
}
\caption{The mean kinetic energy of self-oscillations and the standard
deviation of energy for the system of three self-rotators with mechanical
constraint (\ref{eq19}) with for $\nu = 1.5$ versus the criticality parameter $\mu$}\label{fig7}
\end{figure}

Concerning a number of significant Lyapunov exponents, the same reasoning is
valid as for the previous model. Figure 9 shows a plot of the four Lyapunov
exponents calculated by the Benettin algorithm \cite{33,34,35} depending on the
parameter $\mu $.

\begin{figure}[!t]
\centering{
\includegraphics[width=8cm]{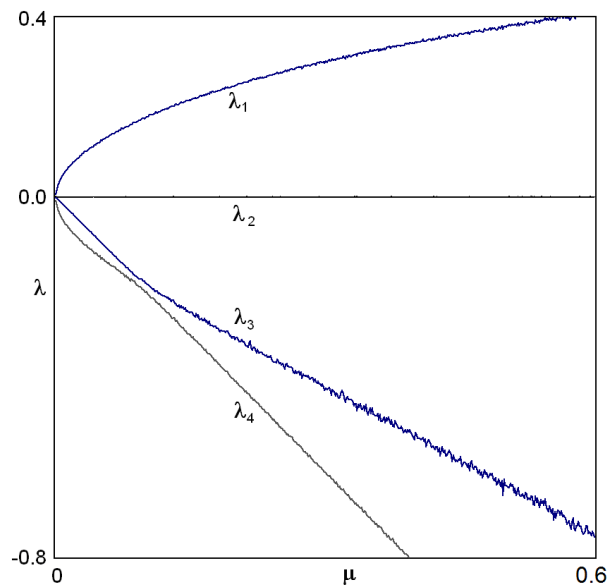}
}
\caption{Lyapunov exponents of the dissipative system of three self-rotators
with mechanical constraint depending on the parameter $\mu $ for $\nu $=1.5.
Two nonphysical exponents violating the condition of mechanical constraint
are excluded}\label{fig8}
\end{figure}

Figure 9 shows histograms for distributions of the angles between stable and
unstable subspaces obtained numerically for the system (\ref{eq16}) with $\nu $ =
1.5 at $\mu $ = 0.06 and 0.39. As seen, the distributions are clearly
distant from zero, the hyperbolicity of the attractors is
confirmed.

\begin{figure}
\centering\includegraphics[width=6cm]{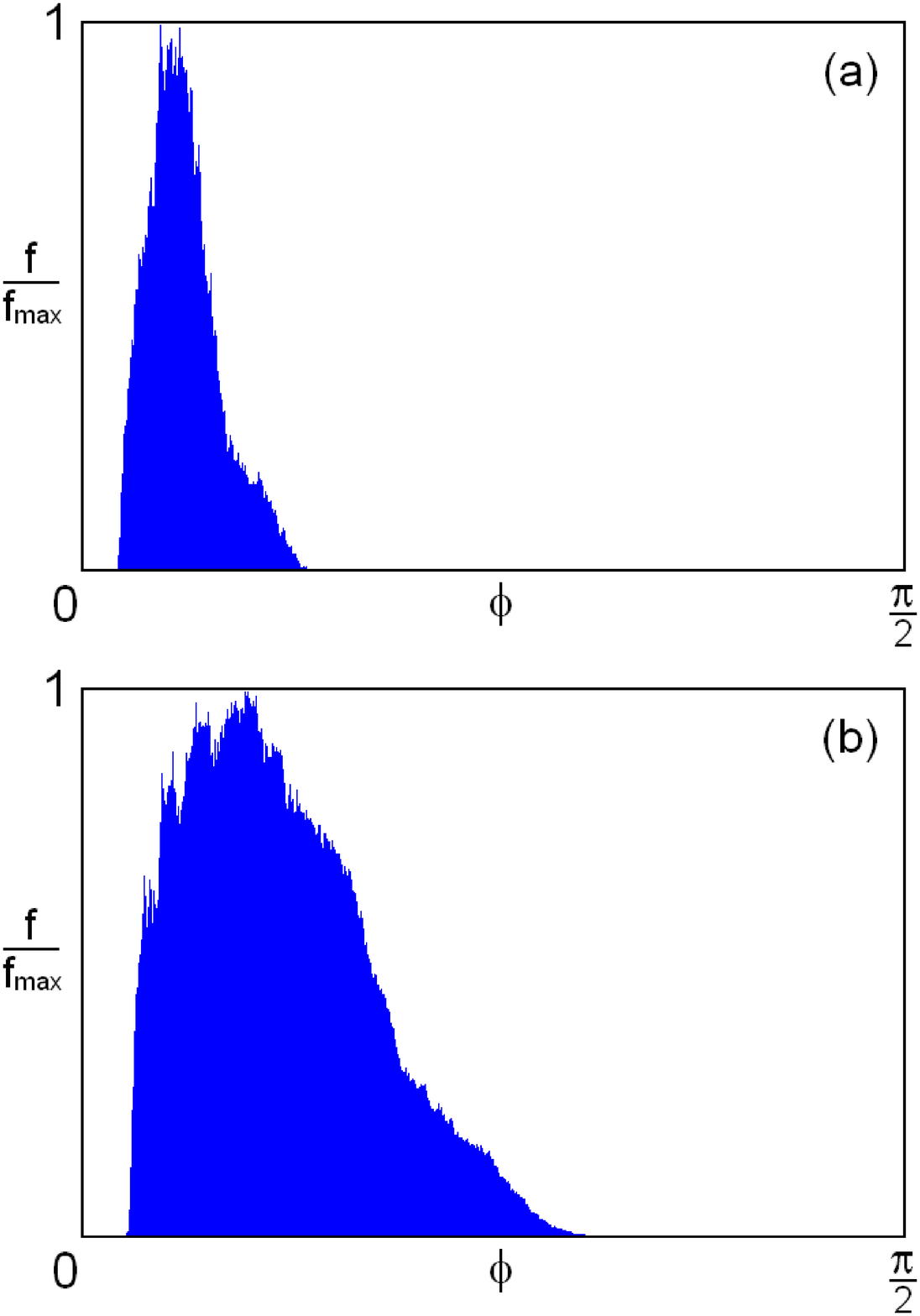}
\caption{Verification of the hyperbolicity criterion for angles in the
system of self-rotators with mechanical constraint for $\mu $=0.06 and $\mu
$=0.39 at $\nu $=1.5}
\end{figure}

\subsection{System of three self-rotators with potential interaction}

From systems with mechanical hinge constraints we turn now to situation
when interaction of the rotators constituting the system is provided
by a potential function depending on the angular variables in such way that
the minimum takes place on the Schwartz surface: $U(\theta _1 ,\theta _2
,\theta _3 ) = \textstyle{1 \over 2}(\cos \theta _1 + \cos \theta _2 + \cos
\theta _3 )^2$. Instead of equations (\ref{eq13}) we write now
\begin{equation}
\label{eq20}
 \ddot {\theta }_i = - \partial U / \partial \theta _i + M_i
 = (\cos \theta _1 + \cos \theta _2 + \cos \theta _3 )\sin \theta _1 + M_i
,\,\,i = 1,2,3,
\end{equation}
where $M_{1,2,3}$ are torques applied to the rotators. Assigning them
according to (\ref{eq15}) we get
\begin{equation}
\label{eq21}
\ddot {\theta }_i = (\mu - \nu \dot {\theta }_i^2 )\dot {\theta }_i - (\cos
\theta _1 + \cos \theta _2 + \cos \theta _3 )\sin \theta _i ,\,\,i = 1,2,3.
\end{equation}

Figure 10 shows the dependencies of the generalized velocities of rotators and
of the kinetic energy on time in the transient process in the system (\ref{eq21}). As a result
of the transient process, a chaotic self-oscillatory regime arises, where
the kinetic energy fluctuates irregularly about a certain mean value.
\begin{figure}
\centering\includegraphics[width=15cm]{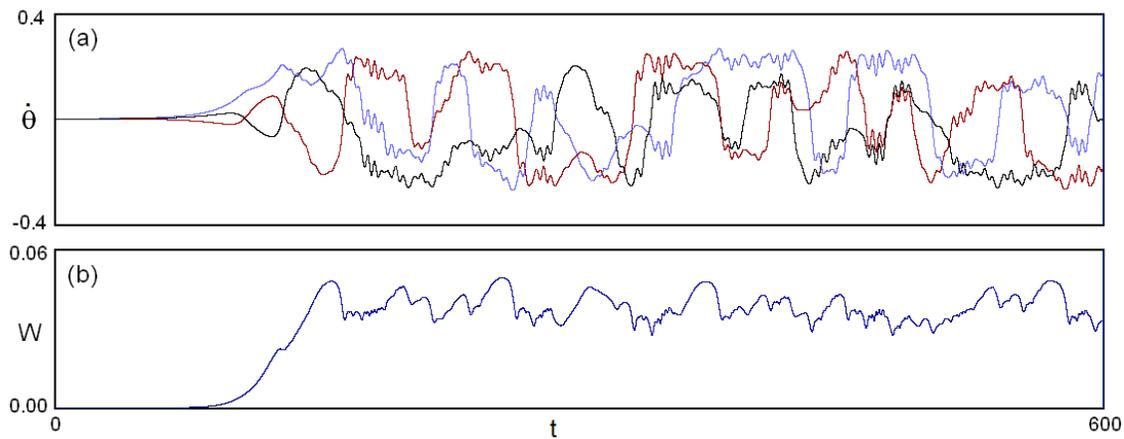}
\caption{Plots for the angular velocities $\dot {\theta }_1 ,\,\,\dot {\theta
}_2 ,\,\,\dot {\theta }_3 $ and kinetic energy $W = \textstyle{1 \over
2}(\dot {\theta }_1^2 + \dot {\theta }_2^2 + \dot {\theta }_3^2 )$ versus
time in the transient process towards chaotic self-oscillations in the
system (\ref{eq21}) at $\mu $=0.06, $\nu $=1.5 (b)}
\end{figure}

Figure 11 illustrates behavior of the trajectories in the configuration
space of the system (\ref{eq21}). At small values of the parameter $\mu $, which
corresponds to a small average energy in the sustained regime, the
trajectories are located near the two-dimensional surface determined by the
equation $\cos \theta _1 + \cos \theta _2 + \cos \theta _3 = 0$
corresponding to the mechanical constraint assumed in the previous
sections. This gives foundation to expect that in this situation the
hyperbolic dynamics are still preserved. However, already here one can
observe that the trajectory is "fluffed" in a direction transversal to the
surface, which reflects the presence of fluctuations in the potential energy
in the course of the motion. This effect, insignificant at small $\mu $,
becomes more pronounced with increase of $\mu $, which can be seen in the
diagram (b). One can expect, and this is confirmed by numerical
calculations, that the nature of the attractor can change due to this
effect, in particular, the dynamics may become non-hyperbolic.
\begin{figure}[!t]
\centering\includegraphics[width=5cm]{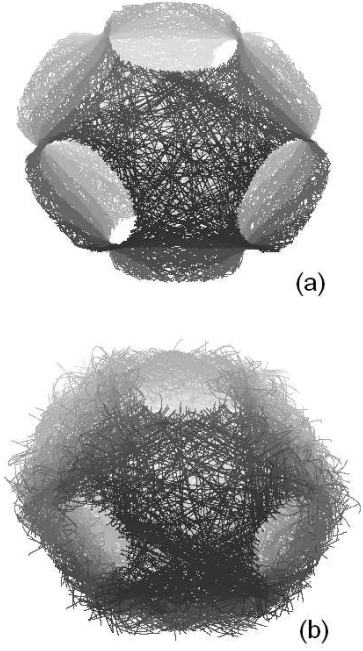}
\caption{Trajectory on the attractor of the system (\ref{eq21}) for $\mu $ = 0.06
(a) and $\mu $ = 0.75 (b) in the three-dimensional configuration space at
$\nu $ = 1.5}
\label{fig11}
\end{figure}

Figure 12 shows the mean kinetic energy of self-oscillations and the
root-mean-square deviation of energy versus parameter $\mu $ for the system
of three self-rotators with potential interaction (\ref{eq21}). In the left part of
the plot the dependences look similar to those shown in Fig. 8, which
indicates preservation of the same type of hyperbolic dynamics. However,
with increase in the parameter, approximately at $\mu  \approx $0.54, one
can see a change in the nature of the dynamics. As calculations show,
instead of chaos in the region, a regular regime occurs.
It corresponds to a limit cycle in the phase space.

\begin{figure}[!t]
\centering\includegraphics[width=6cm]{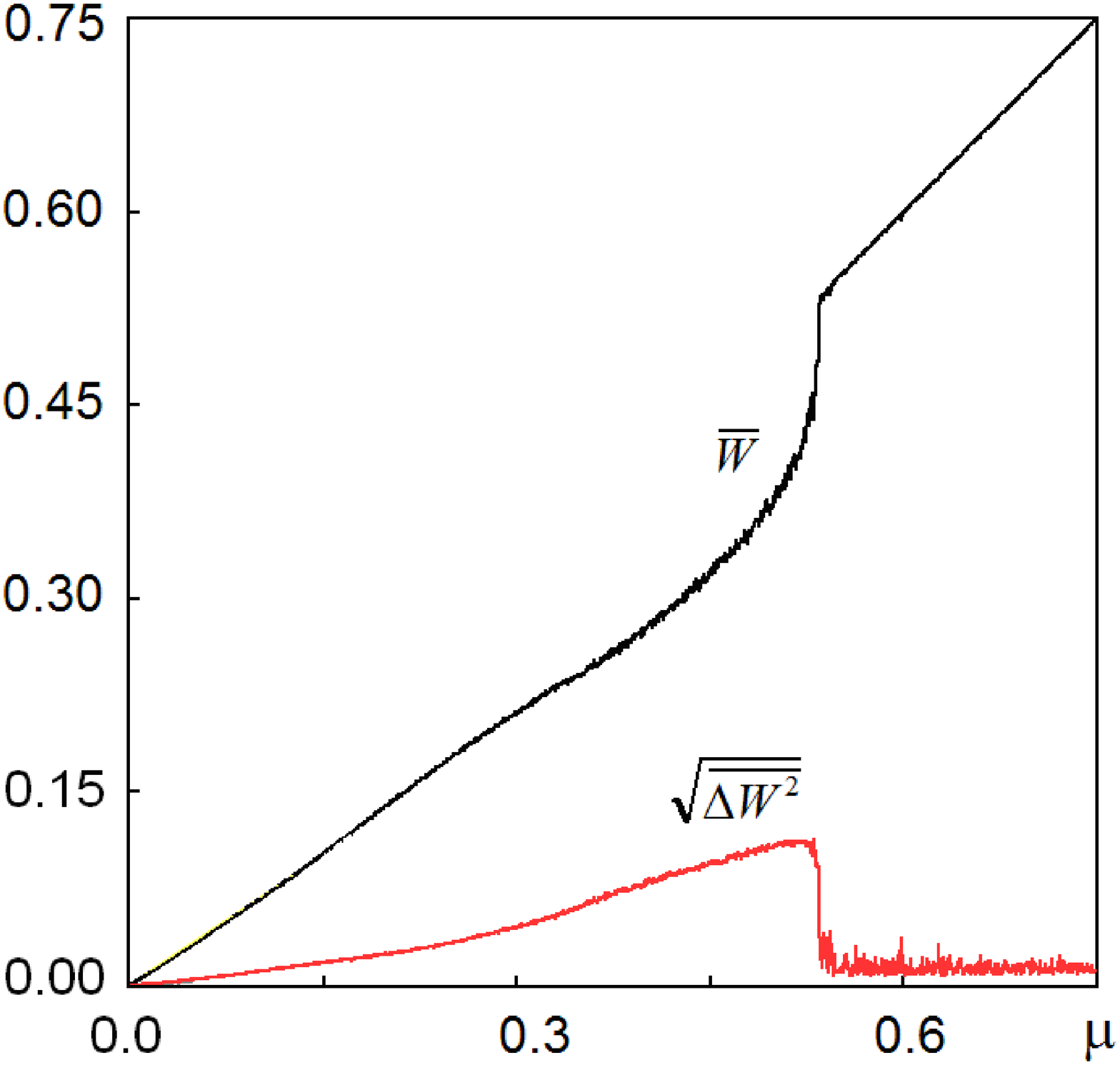}
\caption{Plots of the mean kinetic energy of self-oscillations and the
root-mean-square deviation of energy for a system of three rotators with
potential interaction (\ref{eq21}) for $\nu $ = 1.5}
\label{fig12}
\end{figure}

Figure 13 shows the Lyapunov exponents calculated using the Benettin
algorithm \cite{33,34,35} versus the parameter for the model (\ref{eq21}). For this system,
all six exponents should be taken into account in the analysis as there is
no reason to exclude any of them from the consideration. The presence of a
zero exponent is due to the autonomous nature of the system, and it is
associated with perturbations of a shift along the reference trajectory.

In the region of small $\mu $ we have one positive, one zero, and the other
negative Lyapunov exponents. Dependence of the exponents on the parameter is
smooth here, without irregularities, which suggests that the hyperbolic
chaos persists, as in the original model considered in Section~2. The senior
Lyapunov exponent remains positive, and the dynamics is chaotic up to $\mu
 \approx $0.54. When approaching this point, brokenness arises and
progresses in the graph of the senior exponent that apparently indicates
destruction of the hyperbolicity, though no visible drops to zero with
formation of regularity windows are distinguished. Then, the chaos
disappears sharply, and the system manifests transition to the regular mode,
where the senior exponent is zero, and the others are negative that
corresponds to the limit cycle.
\begin{figure}[!t]
\centering\includegraphics[width=8cm]{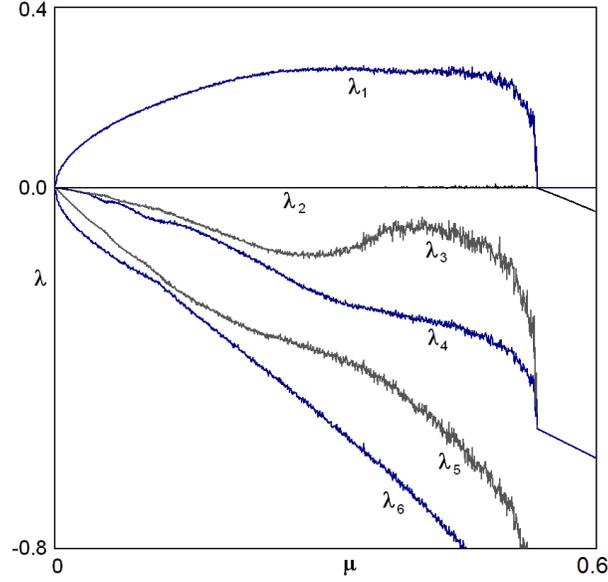}
\caption{Lyapunov exponents of the system of three self-rotators with
potential interaction (\ref{eq21}), depending on the parameter $\mu $ for $\nu
$=1.5}
\label{fig13}
\end{figure}

Figure 14 shows results of testing attractors of the model (\ref{eq21}) using the
criterion of angles. Here the numerically obtained histograms are plotted
for the angles between stable and unstable subspaces of perturbation vectors
for typical chaotic phase trajectories. At small values of $\mu $ the
distributions are disposed at some finite distance from zero angles, i.e.
the test confirms the hyperbolicity. This feature, however, is violated
somewhere in the course of increase of $\mu $. Observe that the histogram
(b) clearly demonstrates the presence of angles close to zero, which
indicates occurrence of tangencies of stable and unstable manifolds, and,
hence, signalizes about non-hyperbolic nature of the attractor. As this
phenomenon was not observed in the models with hinge constraints, it is
natural to assume that absence of hyperbolicity is linked with excursions of
the phase trajectories relating to the attractor outside a narrow neighborhood
of the surface of equal potential in the configuration space.
\begin{figure}[!t]
\centering\includegraphics[width=6cm]{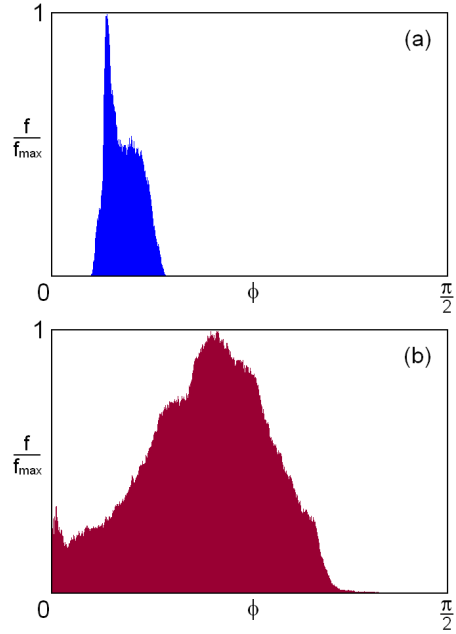}
\caption{Verification of the criterion of angles for three
rotators with potential interaction (\ref{eq21}) at $\nu $=1.5 with
$\mu $=0.06 (a) and $\mu $=0.39 (b). The histogram (a) shows a statistical
distribution separated from zero that indicates the hyperbolic nature of the
attractor. The diagram (b) shows a distribution with positive probability of
angles close to zero indicating occurrences of touches of stable and
unstable manifolds that means violation of the hyperbolicity.}
\label{fig14}
\end{figure}

\subsection{Electronic generator of rough chaos}

To construct an electronic device on the principles discussed in the
previous part of the article, elements analogous to rotators in mechanics
are required. Namely, the state of such an element has to be characterized by a
generalized coordinate defined modulo 2$\pi$ together with its derivative,
the generalized velocity. An appropriate variable of such kind is a phase
shift in the voltage controlled oscillator relative to a reference signal,
like it is practiced in the phase-locked loops \cite{32,33}.

\subsubsection{Circuit diagram of the chaos generator and its functioning}
The circuit shown in Fig. 3 is made up of three similar subsystems
containing voltage-controlled generators, respectively, V1, V2, V3 (in the
diagram they are marked with dashed rectangles). The oscillation phases of
these generators are controlled by the voltages $U_{1}$, $U_{2}$, $U_{3}$
across the capacitors C1, C2, C3. Thus, the voltage outputs vary in time as
$\sin (\omega t + \theta _{1,2,3} )$, where the phases satisfy the equations
\begin{equation}
\label{eq22}
\frac{d\theta _i }{dt} = kU_i ,\,\,\,i = 1,2,3.
\end{equation}
Here $k$ is the steepness factor of frequency tuning of the voltage controlled
generators; Specifically we take $k / 2\pi = 40$ kHz/V. The central
frequency of the generators V1, V2, V3 is 20 kHz, which is provided by the
bias voltage from the source V4. A reference signal with frequency $f = \omega / 2\pi = $20 kHz and amplitude of 1~V is generated by the voltage source V5.
\begin{figure}
\centering\includegraphics[width=15cm]{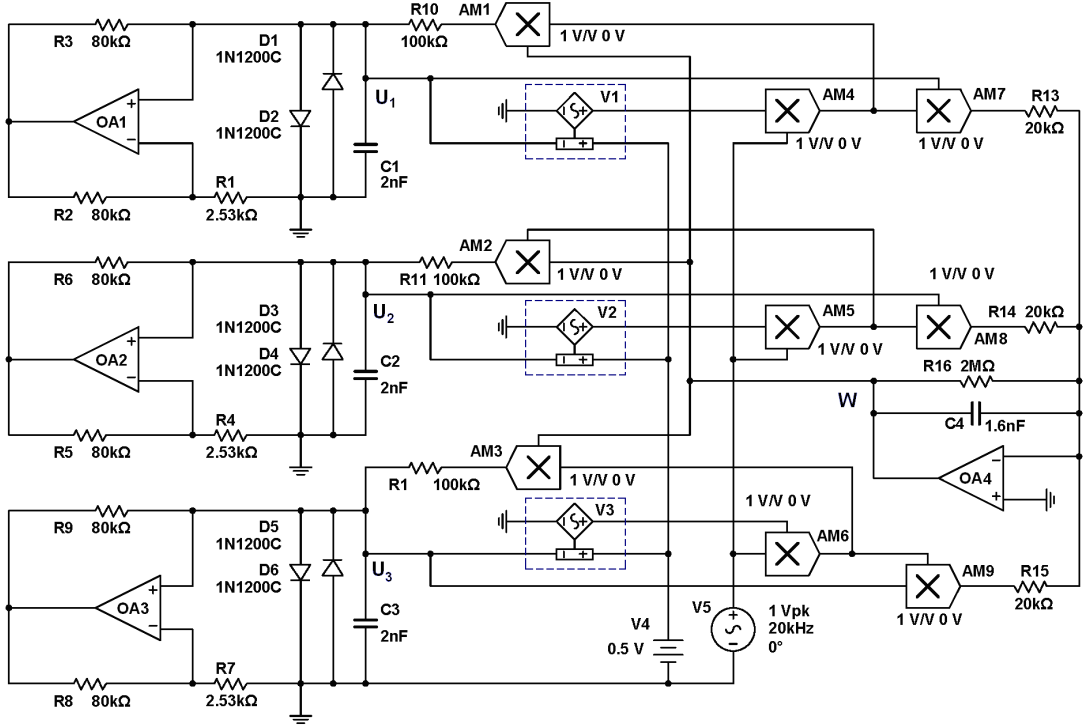}
\caption{Circuit diagram of the chaos generator in the Multisim software
package. Coefficient of frequency control for V1, V2, V3 is $k$/2$\pi $=40
kHz/V}
\label{fig15}
\end{figure}

We write now the Kirchhoff equations for the currents through the capacitors
C1, C2, and C3, assuming that the output voltages of the analog multipliers
AM1, AM2, AM3 are $W_{1,2,3}$. We have
\begin{equation}
\label{eq23}
C\frac{dU_i }{dt} + (R^{ - 1} - g)U_i + \alpha U_i + \beta U_i^3 = \frac{W_i}{R},\,\,i = 1,2,3,
\end{equation}
where $C$ = C1 = C2 = C3 = 2~nF, $R$ = R10 = R11 = R12 = 100~kOhm,
and $I(U) = \alpha U + \beta U^3$ is the characteristic of the nonlinear
element composed of the diodes;
its shape is shown in Fig.16. The equations take into account the negative
conductivity $g = R_2 / R_1 R_3 = R_5 / R_4 R_6 = R_8 / R_7 R_9 $ introduced
by the elements on the operation amplifiers OA1, OA2, OA3. The voltages
$W_{1,2,3}$ are obtained by multiplying the signals $\sin (\omega t + \theta
_{1,2,3} )\cos \omega t$ from AM4, AM5, AM6 by an output signal $W$ of
the inverting summing-integrating element containing the operational
amplifier OA4.
\begin{figure}[!t]
\centering\includegraphics[width=8cm]{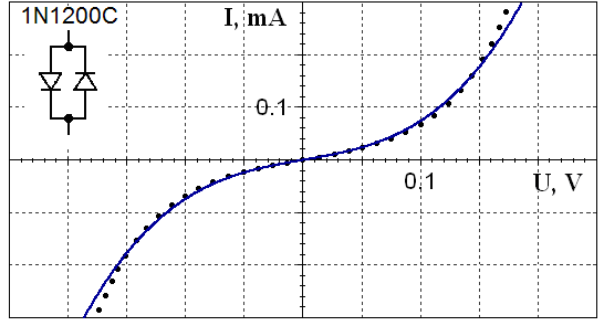}
\caption{Volt-ampere characteristic of a nonlinear element composed of two
parallel diodes 1N1200C. The differential resistance at low voltages is
2.602 kOhm. The points represent the data of Multisim simulating; the black
curve is the approximation: $I(U) \approx \alpha U+\beta U^{3}$=
0.0039$U$+0.035$U^{3}$, where the current is expressed in amperes, and the
voltage in volts}
\label{fig16}
\end{figure}

Input signals for the summing-integrating element are the output voltages of
the multipliers AM7, AM8, AM9, which are expressed as $U_{1,2,3} \sin \omega
t\sin (\omega t + \theta _{1,2,3} )$, so, taking into account the leakage
due to the resistor R16, for the voltage $W$ we can write
\begin{equation}
\label{eq24}
C_0 \frac{dW}{dt} + \frac{W}{r} = - \frac{1}{R_0 }\sum\limits_{i = 1}^3 {U_i
\sin (\omega t + \theta _i )\cos \omega t} ,
\end{equation}
where C$_{0}$=C4=1.6~nF, $R_{0}$ =R13=R14=R15=20 kOhm, $r$=R16=2~MOhm.

Introducing normalized variables
\begin{equation}
\label{eq25}
\tau = \frac{t}{2\sqrt {RCR_0 C_0 } },\,\,\,u_i = 2k\sqrt {RCR_0 C_0 } U_i
,\,\,\,w = 2kR_0 C_0 W
\end{equation}
and parameters
\begin{equation}
\label{eq26}
 \Omega = 2\sqrt {RCR_0 C_0 } \omega ,\,\,\,\mu = 2(gR - \alpha R - 1)\sqrt
{\frac{R_0 C_0 }{RC}} ,\,\,
 \nu = \frac{\beta }{2k^2C\sqrt {RCR_0 C_0 } },\,\,\,\gamma = \frac{2\sqrt
{RCR_0 C_0 } }{rС_0 },
\end{equation}
we obtain the equations
\begin{equation}
\label{eq27}
\begin{array}{l}
 \dot {\theta }_i = u_i ,\,\,\, \\
 \dot {u}_i = \mu u_i - \nu u_i^3 + 2w\sin (\Omega \tau + \theta _i )\cos
\Omega \tau ,\,\,\,\,i = 1,\,\,2,\,\,3, \\
 \dot {w} = - \gamma w - 2\sum\limits_{i = 1}^3 {u_i \sin (\Omega \tau +
\theta _i )\cos \Omega \tau } , \\
 \end{array}
\end{equation}
where the dot means the derivative over the dimensionless time $\tau$.

Taking into account that $\Omega \gg 1$ one can simplify the equations
assuming that $u_{i}$ and $w$ vary slowly on the high-frequency period. Namely,
we perform averaging in the right-hand parts setting
\begin{equation}
\label{eq28}
 \overline {\sin (\Omega \tau + \theta _i )\cos \Omega \tau }
 = \overline {\cos ^2\Omega \tau \sin \theta _i + \sin \Omega \tau \cos
\Omega \tau \cos \theta _i }
 = \textstyle{1 \over 2}\sin \theta _i
\end{equation}
and arrive at the equations
\begin{equation}
\label{eq29}
\begin{array}{l}
 \dot {\theta }_i = u_i ,\,\,\,\dot {u}_i = \mu u_i - \nu u_i^3 + w\sin
\theta _i ,\,\, i = 1,\,\,2,\,\,3, \\
 \dot {w} = - \gamma w -
\sum\limits_{i = 1}^3 {u_i \sin \theta _i } . \\
 \end{array}
\end{equation}

Finally, supposing $\gamma \ll 1$ we can neglect the term $\gamma w$ in the
last equation and to integrate it with substitution of $u_{1,2,3}$ from the
first equation. Then we obtain $w \approx \cos \theta _1 + \cos \theta _2 +
\cos \theta _3 $, and the final result corresponds exactly to the equations
(\ref{eq21}).

\subsubsection{Circuit simulation}

Figure 17 shows screenshots of the signals $U_{1,2,3}$ copied from the
virtual oscilloscope screen when simulating the dynamics of the circuit in
Multisim.\footnote{ When modeling in the Multisim environment, there is a
problem of launch of the system due to a long time it takes to escape from
the trivial state of equilibrium. The waveforms shown in Fig.~17 refer to
the dynamics on the attractor, the transient process is excluded.} Visually,
the signals look chaotic, without any apparent repetition of forms. Figure~18 shows the spectrum of the signal $U_{1}$, obtained with the help of a virtual
spectrum analyzer. Continuous spectrum, as it should be for a chaotic
process, is characterized by slow decrease of the spectral density with
frequency and is of rather good quality in the sense of lack of pronounced
peaks and dips. Because of the symmetry of the circuit, all three time
dependences for voltages $U_{1,2,3}$ are statistically equivalent, and their
spectra as verified have the same form.
\begin{figure}[!t]
\centering\includegraphics[width=15cm]{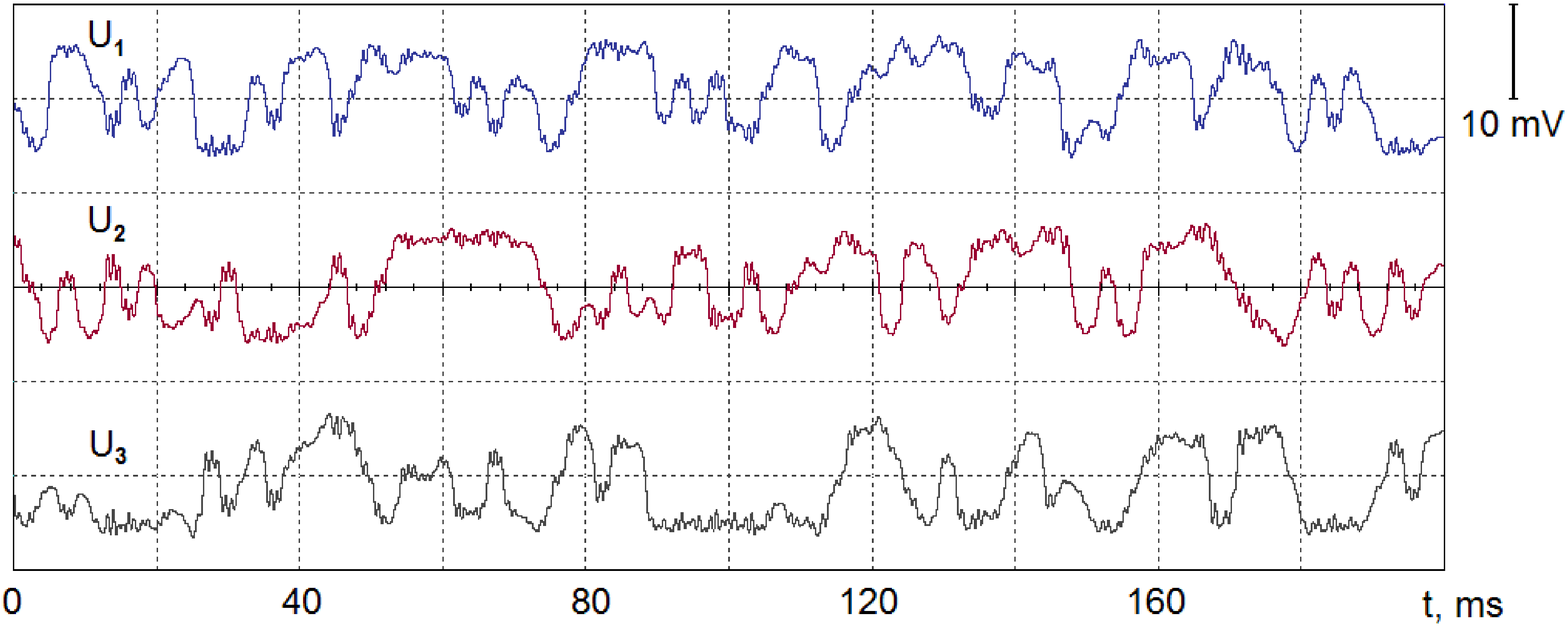}
\caption{Voltages on capacitors C1, C2 and C3 obtained from the virtual
oscilloscope screen when modeling the circuit in the Multisim environment.
The scale along the vertical axis is indicated at the top in the right part
of the figure}
\label{fig17}
\end{figure}[!t]
\begin{figure}
\centering\includegraphics[width=6cm]{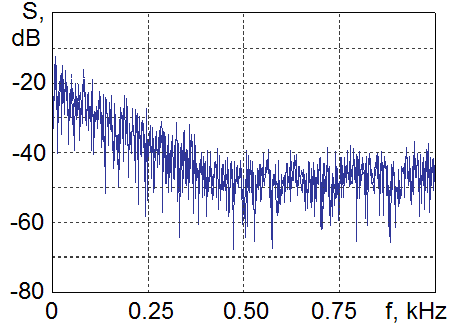}
\caption{The power spectrum for the signal $U_{1}$ obtained as a snapshot from
the virtual spectrum analyzer screen when modeling the dynamics of the
circuit in the Multisim environment}
\label{fig18}
\end{figure}

To monitor the angular variables $\theta _{1,2,3}$ the circuit was
supplemented by three special signal processing modules, in each of which
the signal from the output of the voltage-controlled oscillators V1, V2, V3
was multiplied by the reference signals $\sin \omega t$ and $\cos \omega t$
(Fig.~19). After filtering with extraction of the low-frequency components,
the obtained three pairs of signals ($x_{k}$, $y_{k})$, $k$=1,2,3, are fed to
inputs of three oscilloscopes, and during the circuit operation, these
signals are recorded into a file in computer for further processing with
calculation of $\theta _k = \arg (x_k + iy_k )$, $k$=1,2,3.
\begin{figure}[!t]
\centering\includegraphics[width=8cm]{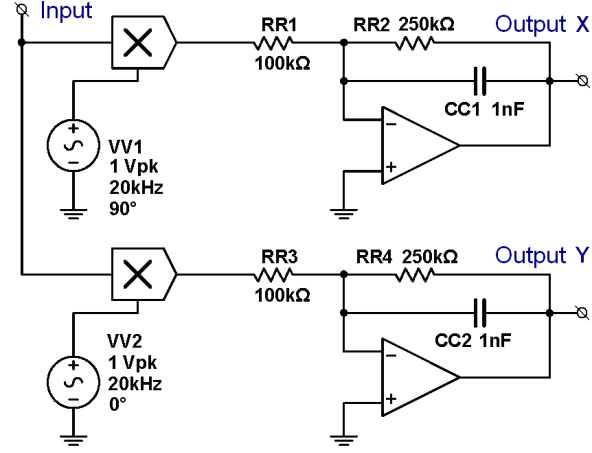}
\caption{Signal processing module for constructing phase
trajectories in the configuration space in circuit simulation. The circuit
in Figure 15 was supplemented by three such modules, for which the input
signals were time-varying voltages from the output of generators V1, V2, V3.
It is essential that the alternating voltage sources VV1 and VV2 are phase
shifted by 90 \r{ } relative to each other}
\label{fig19}
\end{figure}

\subsubsection{Dynamics of the chaos generator: numerics}

In a framework of circuit simulation it is difficult to extract certain
characteristics, for example, Lyapunov exponents, and it is not possible to
verify the hyperbolic nature of chaos. Therefore, we turn to comparison of
the results of numerical integration of the model equations (\ref{eq27}), (\ref{eq29}), and
(\ref{eq21}), for which the corresponding analysis can be performed in computations.

Using the nominal values of the circuit components in Fig. 16 and the
formulas of the previous section, we find the parameters appearing in
equations (\ref{eq27}), (\ref{eq29}) and (\ref{eq21}):
\begin{equation}
\label{eq30}
\mu = 0.07497,\,\,\nu = 1.73156,\,\,\gamma = 0.05,\,\,\,\Omega = 20.1062.
\end{equation}

\begin{figure}[!b]
\centering\includegraphics[width=15cm]{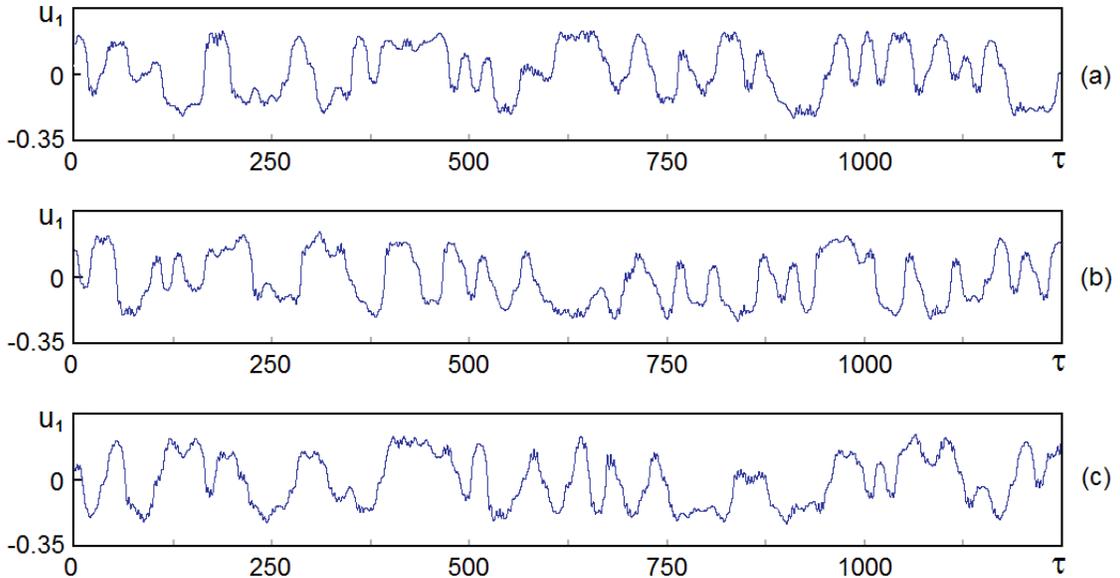}
\caption{The time dependences of the variable $u_{1}$ obtained in the
numerical solution of the equations for the models (\ref{eq27}), (\ref{eq29}) and (\ref{eq21}),
respectively, (a), (b) and (c)}
\label{fig20}
\end{figure}
Figure 20 shows plots for the dimensionless variable $u_{1}$ versus the
dimensionless time obtained from the numerical integration of the equations
(\ref{eq27}) - panel (a), (\ref{eq29}) - panel (b) and (\ref{eq21}) - panel (c). Although there is
no exact coincidence of the graphs on the diagrams (a), (b), (c) (because of the
chaotic nature of the dynamics and its sensitivity to variations of
the initial conditions), they are in reasonable agreement (note general view of
the realizations and characteristic scales along the coordinate axes).
This can be regarded as a confirmation of the validity of the approximations
made in the course of sequential simplification of the model. This kind of
correspondence can also be observed when comparing the graphs to the
oscillograms in Fig. 17, obtained in the Multisim simulation.

Figure 21 shows power spectrum generated by the model (\ref{eq27}), which is in
reasonsble agreement with that obtained by the circuit simulation (Fig. 19),
and with that of the system (\ref{eq8}), (\ref{eq9}) (Fig. 2).

\begin{figure}[!t]
\centering\includegraphics[width=8cm]{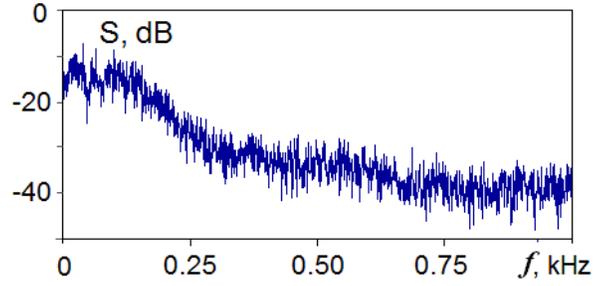}
\caption{The power spectrum for the signal generated by the time dependence
of the variable u1 in the model (\ref{eq27}) with the parameters (\ref{eq30})}
\label{fig21}
\end{figure}

As one can verify, the dynamics of the electronic device proceeds in such a
way that the trajectory in the space of coordinate variables ($\theta
_{1}$,$\theta _{2}$,$\theta _{3})$ is located near the Schwarz
surface. It is illustrated in Figure 22, where the trajectories obtained by
numerical integration of the equations of the models (\ref{eq27}), (\ref{eq21}) and (\ref{eq21}) are
shown, respectively, in panels (a), (b), (c). They can be compared with
Figure~1 for the original geodesic flow on the surface of negative
curvature. It can be seen that the trajectories are close to the Schwartz
surface, although they are not exactly located on it: they are slightly
"fluffy" in the transverse direction. This effect becomes more pronounced
with increase of the parameter $\mu $.

\begin{figure}[!t]
\centering\includegraphics[width=10cm]{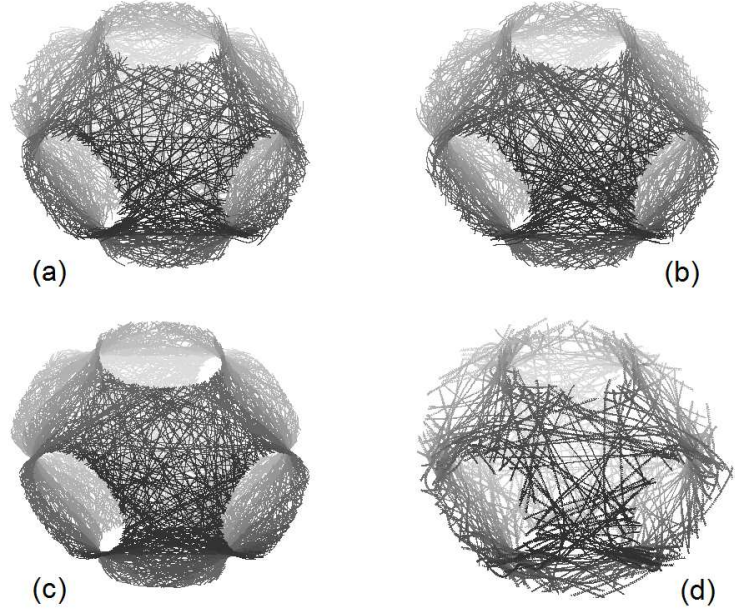}
\caption{Trajectories in the three-dimensional space $(\theta _1 ,\,\theta _2
,\,\theta _3 )$ for model systems described by equations (\ref{eq27}) (a), (\ref{eq29}) (b),
and (\ref{eq21}) (c), and for the electronic device simulated in Multisim
corresponding to the scheme in Fig. 15 with the indicated nominal of the
components (d)}
\label{fig22}
\end{figure}

The diagram (d) is obtained by processing data of circuit
simulation in Multisim environment using the modification of the scheme
described at the end of subsection 5.2. To construct the diagram by
processing the recorded data, at each instant of time, three angular
variables $\theta _k = \arg (x_k + iy_k )$, $k$=1,2,3, were computed and a
corresponding point was displayed on the graph. The diagram clearly
demonstrates that the functioning of the device corresponds to dynamics
on trajectories near the Schwarz surface, as well as in the models (\ref{eq27}),
(\ref{eq21}), and (\ref{eq21}).
\begin{figure}[!t]
\centering\includegraphics[width=8cm]{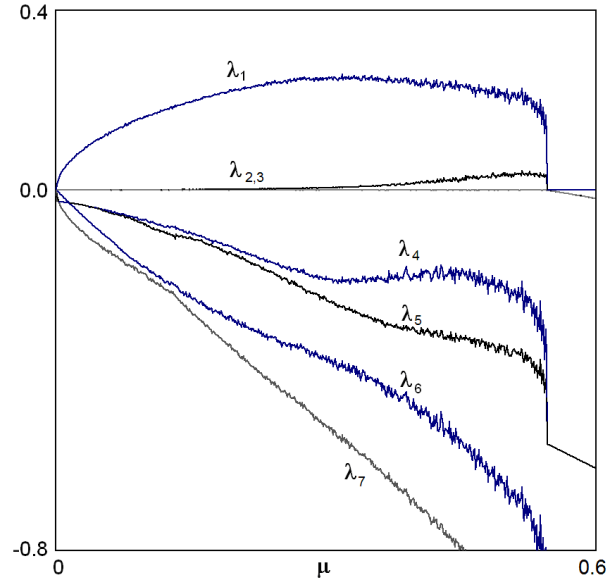}
\caption{Lyapunov exponents of the system (\ref{eq27}) depend on the parameter $\mu $
for the remaining parameters given in accordance with (\ref{eq20})}
\label{fig23}
\end{figure}
\begin{figure}[!t]
\centering\includegraphics[width=15cm]{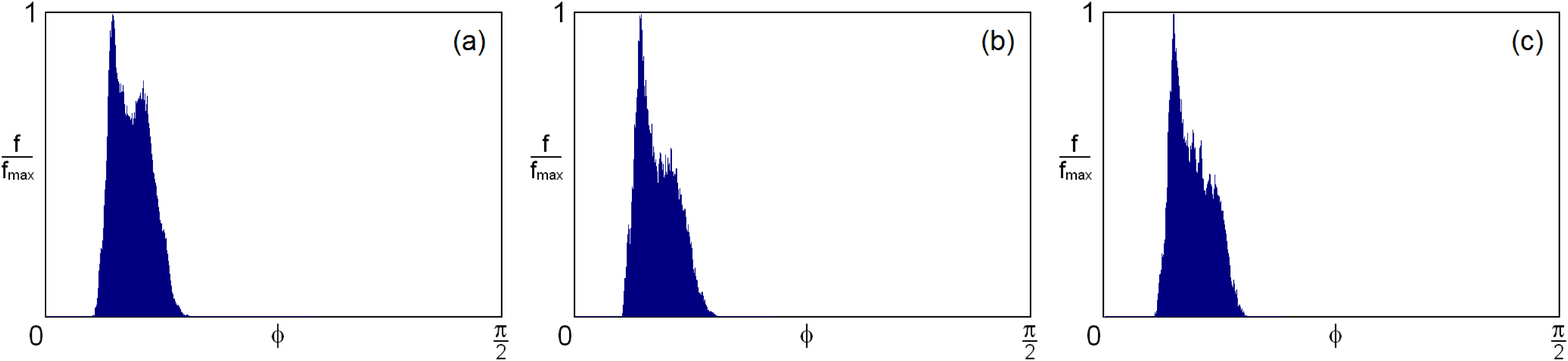}
\centering\includegraphics[width=15cm]{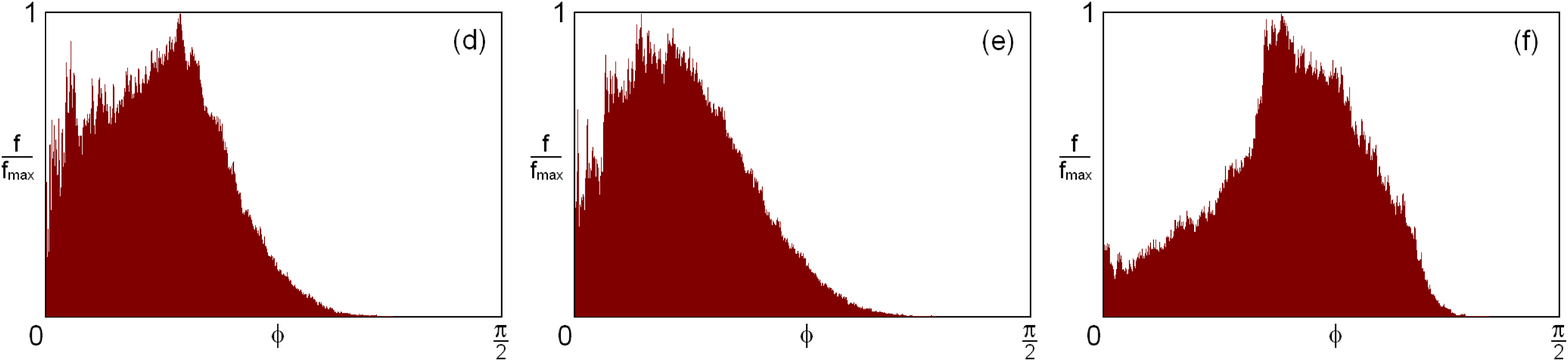}
\caption{The histograms of the distribution of angles between stable and
unstable subspaces obtained numerically: (a) in model (\ref{eq27}), (b) in model
(\ref{eq29}) and (c) in model (\ref{eq21}) for R$_{1,4,7}$=2.53 k$\Omega $, $\mu $=0.07497,
when the distributions are distant from zero angles, and the attractor is
hyperbolic. The bottom row represents histograms for the case of absence of
hyperbolicity in model (\ref{eq27}) (d), in model (\ref{eq29}) (e) and in model (\ref{eq21}) (f) for R$_{1,4,7}$=2.5 k$\Omega $, $\mu $=0.4544, when the distributions manifest
presence of angles around zero.}
\label{fig24}
\end{figure}
Figure 23 shows a graph of seven Lyapunov exponents calculated for the model
(\ref{eq27}) using the Benettin algorithm \cite{33,34,35}. In the entire range of the
parameter $\mu $, we have one positive, two close to zero, and the other
negative Lyapunov exponents. The dependence of the Lyapunov exponents on the
parameter in this region is smooth, without peaks and dips, which allows us
to assume that the hyperbolic nature of chaos retains.

In Fig. 24 histograms of the distributions of the angles obtained for the
attractors of the systems (\ref{eq27}), (\ref{eq21}) and (\ref{eq21}) are shown.
The upper row corresponds to the parameters of the circuit
in Fig. 15 (see (\ref{eq30})). For all three models the
diagrams look similar, and the distributions are clearly separated from zero.
Thus, the test confirms hyperbolicity of the attractor.
For comparison, histograms obtained in the situation when
hyperbolicity is violated are presented in the lower row, at
sufficiently large $\mu $. In fact, they demonstrate a
statistically significant presence of angles near zero, which indicates
occurrence of tangencies of stable and unstable manifolds and nonhyperbolic
nature of the attractor. Apparently, disappearance of the hyperbolicity
occurs due to the deviating trajectories on the attractor from the Schwarz
surface.

\section*{Conclusion}

Hyperbolic chaos, which in dissipative systems corresponds to hyperbolic
attractors, is characterized by
roughness, or structural stability, as a mathematically rigorous attribute.
Therefore, devices that generate such chaos must be preferred for any practical
applications because of low sensitivity to variation of parameters,
imperfections, disturbances, etc.

A reasonable approach to constructing systems with hyperbolic chaos is to
start with a formal mathematical example, where such chaos takes
place, and modify it by variations of functions and parameters in the
respective equations. For small variations, the hyperbolic nature of the
dynamics retains, as it is ensured by structural stability. However, as the
detachment from the original system increases, violation of hyperbolicity
becomes possible, and it should be monitored using quantitative criteria,
one of which is analysis of the distribution of angles between stable and
unstable manifolds of relevant trajectories.

In this paper we have considered several variants of systems with chaotic
dynamics inspired by the problem of geodesic flow on a surface of negative
curvature. Mechanical systems are based on the Thurston -- Weeks -- Hunt -
MacKay hinge mechanism made up of three rotating disks. Their motions are either
mechanically constrained with hinges and rods, or the constraint is replaced by potential interaction.

By analogy with the mechanical systems, a construction of electronic generator of rough chaos is proposed, electronic analog circuit is designed, corresponding equations are derived and computer
study of the chaotic dynamics is carried out. The device
was also modeled using the Multisim environment.

In contrast to the previously considered electronic circuits with hyperbolic
attractors \cite{40, 43,44,45,46,47,48}, in this case hyperbolicity is characterized by an
approximate uniformity in the expansion and compression of the phase volume
elements in the course of evolution in continuous time. Due to this, the generator possesses rather good spectral properties providing a smooth distribution of the
spectral power, without peaks and dips.

Although the specific scheme described in the article operates in
a low-frequency range (kHz), it seems possible to build similar devices also
in higher frequency bands.

\section*{Acknowledgement}
The work was supported by Russian Science Foundation, grant No 15-12-20035 (sections A, B, C -- model of the geodesic flow on the Schwarz surface and mechanical self-oscillating model systems) and, partially, by RFBR grant No 16-02-00135 (section D -- circuit design and analysis of the electronic device).

\end{document}